\documentstyle[prd,aps,preprint,tighten,floats]{revtex}
\begin{document}
\draft
\preprint{UPR-650-T}
\date{March 1995}
\title {Addendum to Supersymmetric Dyonic Black Holes in Kaluza-Klein Theory}
\author {Mirjam Cveti\v c
\thanks{E-mail address: cvetic@cvetic.hep.upenn.edu}
and Donam Youm\thanks{E-mail address: youm@cvetic.hep.upenn.edu}}
\address {Physics Department \\
          University of Pennsylvania, Philadelphia PA 19104-6396}
\maketitle
\begin{abstract}
{We complete the study of 4-dimensional (4-d), static, spherically
symmetric, supersymmetric black holes (BH's) in Abelian $(4+n)$-d
Kaluza-Klein theory, by showing that for such solutions $n$ electric
charges $\vec{\cal Q} \equiv (Q_1,...,Q_n)$ and $n$ magnetic charges
$\vec{\cal P} \equiv (P_1,...,P_n)$ are subject to the constraint
$\vec{\cal P}\cdot \vec{\cal Q}=0$.  All such solutions can be obtained
by performing the $SO(n)$ rotations, which do not affect the 4-d
space-time metric and the volume of the internal space, on the
supersymmetric $U(1)_M\times U(1)_E$ BH's, {\it i.e.}, supersymmetric
BH's with a diagonal internal metric.}
\end{abstract}
\vskip 1.5cm

In Ref. \cite{CY}, 4-dimensional (4-d), static, spherically symmetric,
supersymmetric black holes (BH's) in Abelian ($4+n$)-d Kaluza-Klein
(KK) theory were obtained with a diagonal internal metric Ansatz.
Such BH's correspond to $U(1)_M\times U(1)_E$ configurations,
{\it i.e.}, they have at most one electric charge $Q$
and one magnetic charge $P$ which necessarily arise from different
$U(1)$ groups.  Here, we complete the study of 4-d static, spherically
symmetric, supersymmetric BH's in Abelian KK  theories by addressing
the corresponding solutions with a general non-diagonal internal
metric Ansatz.  We show that $n$ electric charges $\vec{\cal Q}
\equiv (Q_1,...,Q_n)$ and $n$ magnetic charges $\vec{\cal P}
\equiv (P_1,...,P_n)$ of such BH's are subject to the constraint
$\vec{\cal P}\cdot \vec{\cal Q}=0$.  All such solutions can be
obtained by performing the global $SO(n)$ rotations\cite{CYI} on
the supersymmetric $U(1)_M\times U(1)_E$ BH solutions.

Throughout we use the notation specified in Ref. \cite{CY}.

As discussed in Section 4.1 of Ref. \cite{CY}, the supersymmetric BH
configurations are solutions of the Killing spinor equations,
which correspond to the vanishing of supersymmetric transformations
on the dimensionally reduced $(4+n)$-d gravitino, {\it i.e.},
$\delta \psi_{\mu}^{\bf m} =\delta \psi_{\tilde{\mu}}^{\bf m}=0$.
Note, $\psi_\mu^{\bf m}$ and $\psi_{\tilde \mu}^{\bf m}$
(${\bf m}=1,\cdots, 2^{[{n\over2}]},\ \mu={0,\cdots,3},\
{\tilde \mu}=4,\cdots , (n+3)$) are the corresponding 4-d
gravitino(s) and dilatino(s), obtained by the dimensional reduction
of $(4+n)$-d gravitino(s).  (See Eqs. (3.6) and (3.7) of Ref. \cite{CY}
for the explicit form of the corresponding supersymmetry transformations.)

For spherically symmetric configurations with the 4-d metric Ansatz
given by Eq. (4.2) of Ref. \cite{CY} and, however, now with a
general non-diagonal internal metric Ansatz, one can rewrite
the $t$-component $\delta\psi_{t}^{\bf m}=0$ of the 4-d gravitino Killing
spinor equation, analogous to Eq. (4.9) in Ref. \cite{CY}, as:
\begin{equation}
{R\over{ \sqrt \lambda}}\left(\partial_r\lambda - {1 \over \alpha}
\lambda \partial_r \varphi\right)\gamma^{03}\varepsilon -
\sum_{\tilde a=4}^{n+3}\tilde{\bf Q}^{\tilde a}(\gamma^{35} \otimes
\gamma^{\tilde a})\varepsilon = 0 \ ,
\label{time}
\end{equation}
where
$\tilde{\bf Q}^{\tilde a}\equiv {\rm e}^{-{\alpha\over 2}\varphi}
(\Phi^{-1})^{\tilde a}_{\tilde \pi} Q^{\tilde \pi}$.  Note,
$\alpha=\sqrt{{n+2}\over n}$, ${\Phi}_{\tilde \pi}^{\tilde a}$
(${\tilde a},\tilde \pi= 4, \cdots , n+3$) is the $n$-bein of the
unimodular part of the internal metric, $\varphi$ is the dilaton
(the volume of the internal space), and $\lambda$ and $R$ are the
$(t,t)$ and $(\theta,\theta)$ components of the 4-d metric $g_{\mu\nu}$,
respectively.  Here, $Q^{\tilde\pi}$ are electric charges obtained by
integrating the Maxwell's equation $\partial_r({\rm e}^{\alpha \varphi}
R\lambda {\Phi}_{\tilde \pi}^{\tilde a}{\Phi}_{\tilde \lambda}^{\tilde
a}F^{\tilde{\lambda}\ tr})=0 $.

The $\theta$-component $\delta\psi_{\theta}^{\bf m}=0$ of the 4-d
gravitino Killing spinor equation, analogous to Eq. (4.10),
supplemented by the constraint Eq. (5.2) in Ref. \cite{CY},
which is always satisfied by the Killing spinors of spherically
symmetric configurations, assumes the following form:
\begin{equation}
\left[2\sqrt{R}-\sqrt{\lambda}\left({{\partial_rR}}
- {1 \over \alpha}R\partial_r \varphi\right)\right]
\gamma^{13} \varepsilon -\sum_{\tilde a=4}^{n+3}{\tilde{\bf  P}}^{\tilde a}
(\gamma^{25} \otimes \gamma^{\tilde a})\varepsilon = 0 \ ,
\label{thet}
\end{equation}
where ${\tilde{\bf P}}^{\tilde a}= {\rm e}^{{\alpha\over 2}\varphi}
\Phi^{\tilde a}_{\tilde \pi}P^{\tilde \pi}$.  Here, $P^{\tilde \pi}$ are
magnetic charges satisfying the Maxwell's equation $\partial_\theta
({\rm e}^{\alpha \varphi}R\lambda \sin\theta \Phi_{\tilde \pi}^{\tilde a}
{\Phi}_{\tilde \lambda}^{\tilde a} F^{\tilde{\lambda}\ \theta\phi})=0$ with
$F^{\tilde{\lambda}}_{\theta\phi}=P^{\tilde{\lambda}}\sin\theta$.

Eqs. (\ref{time}) and (\ref{thet})  can be satisfied if and only if
the lower two-component spinors $\varepsilon^{\bf m}_\ell$ and
the upper two-component spinors $\varepsilon^{\bf m}_u$
($(\varepsilon^{\bf{m}})^T\equiv(\varepsilon^{\bf{m}}_u ,
\varepsilon^{\bf{m}} _\ell)$) satisfy the  following constraint:
\begin{equation}
\eta_Q \sum_{\tilde a=4}^{n+3}
\tilde{\bf Q}^{\prime\ \tilde a} (\gamma^{\tilde a})^{\bf{m}}_{\bf{n}} \
\varepsilon^{\bf{n}}_\ell=\varepsilon^{\bf{m}}_u=
i\eta_P \sum_{\tilde a=4}^{n+3}\tilde{\bf
P}^{\prime\ \tilde a} (\gamma^{\tilde a})^{\bf{m}}_{\bf{n}} \
\varepsilon^{\bf{n}}_\ell \ , \ \ \eta_{Q,P}=\pm 1\ ,
\label{SC}
\end{equation}
where $i\equiv \sqrt{-1}$.  Here, ${{\tilde{\bf Q}}}^{\prime\ \tilde{a}}
\equiv {\tilde{\bf Q}}^{\tilde a}/[R\lambda^{-1/2}(\partial_r\lambda-
{1\over\alpha}\lambda\partial_r\varphi)]$ and
${{\tilde{\bf P}}}^{\prime\ \tilde{a}}\equiv{\tilde{\bf P}}^{\tilde
a}/[2\sqrt{R}-\sqrt{\lambda}(\partial_rR-{1\over\alpha}
R\lambda\partial_r\varphi)]$ satisfy the constraint:
\begin{equation}
\sum_{{\tilde a}=4}^{n+3}(\tilde{\bf Q}^{\prime\ \tilde{a}})^2
=\sum_{{\tilde a}=4}^{n+3}(\tilde{\bf P}^{\prime\ \tilde{a}})^2=1\ .
\label{QP}
\end{equation}
Multiplying the left-most hand side of Eq. (\ref{SC}) by
$\sum_{\tilde b=4}^{n+3}\tilde{\bf P}^{\prime \ \tilde{b}}
(\gamma^{\tilde b})$ and the right-most hand side by
$\sum_{\tilde b=4}^{n+3}\tilde{\bf Q}^{\prime\ \tilde b}(\gamma^{\tilde
b})$, and summing the two resultant equations, along with the identity
$\{\gamma^{\tilde a},\gamma^{\tilde b}\}=-2\delta^{\tilde a\tilde b}$,
one has the result:
\begin{equation}
\sum_{{\tilde a}=4}^{n+3}\tilde{\bf P}^{\prime\ \tilde a}
\tilde{\bf Q}^{\prime\ \tilde a}=0 \ , \label{PQZ}
\end{equation}
or equivalently:
\begin{equation}
\sum_{{\tilde \pi}=4}^{n+3}P^{\tilde \pi}
Q^{\tilde \pi}= \vec{\cal P}\cdot\vec{\cal Q}=0\ .\label{PQZZ}
\end{equation}
Therefore, supersymmetric BH's have constrained charge configurations
with the electric charge vector $\vec{\cal Q}$ and the magnetic charge
vector $\vec{\cal P}$ orthogonal to one another.

New supersymmetric BH solutions can be generated by performing the
global $SO(n)$ rotations on known supersymmetric BH solutions.
Namely, the effective 4-d KK Lagrangian density (see Eq. (3.4)
of Ref. \cite{CY}) is invariant under the global $SO(n)$
transformations \cite{CYI}:
\begin{equation}
{\Phi}^{\tilde \lambda}_{\tilde a}\rightarrow
U^{\tilde \lambda}_{\tilde \pi}{\Phi}^{\tilde \pi}_{\tilde a}\ ,
\ \ \ \  A_{\lambda}^{\tilde \lambda}\rightarrow
U^{\tilde \lambda}_{\tilde \pi}A_{\lambda}^{\tilde \pi} \ ,
\label{SON}
\end{equation}
where $U$ is an $n\times n$ matrix of the $SO(n)$ transformations.
Transformations (\ref{SON}) affect the unimodular part of the
internal metric and the gauge fields, however, they leave the
4-d space-time metric $g_{\mu\nu}$ and the dilaton $\varphi$ intact.
In addition, since the $SO(n)$ transformations are also the
symmetry of the Killing spinor equations, the transformed
solutions remain supersymmetric, {\it i.e.}, they also satisfy the
Killing spinor equations.

The $SO(n)$ transformations (\ref{SON}) on the $U(1)_M\times U(1)_E$
supersymmetric BH's generate solutions with $n$ electric
$\vec{\cal Q}$ and $n$ magnetic $\vec{\cal P}$ charges which are
subject to the constraint $\vec{\cal P}\cdot \vec{\cal Q}=0$
\cite{CYI}.  Therefore, those are all the 4-d static, spherically
symmetric, supersymmetric BH's in Abelian KK theories.

\acknowledgments
The work is supported by U.S. DOE Grant No. DOE-EY-76-02-3071,
and the NATO collaborative research grant CGR 940870.

\end{document}